\documentclass[prb,showpacs,superscriptaddress,twocolumn,amssymb,amsfonts]{revtex4-1}
\usepackage{amsmath}
\usepackage{bm}
\usepackage{epsfig}
\usepackage{graphics}
\usepackage{dsfont}

\def\beq{\begin{equation}}
\def\eeq{\end{equation}}
\def\beqn{\begin{eqnarray}}
\def\eeqn{\end{eqnarray}}

\usepackage{color}

\begin{document}
\title{Conflicting Symmetries in Topologically Ordered Surface States of Three-dimensional Bosonic Symmetry Protected Topological Phases}
\author{Gil Young Cho}
\affiliation{Department of Physics, Institute for Condensed Matter Theory, University of Illinois, 1110 W. Green St., Urbana IL 61801-3080, U.S.A.}
\author{Jeffrey C. Y. Teo}
\affiliation{Department of Physics, Institute for Condensed Matter Theory, University of Illinois, 1110 W. Green St., Urbana IL 61801-3080, U.S.A.}
\author{Shinsei Ryu}
\affiliation{Department of Physics, Institute for Condensed Matter Theory, University of Illinois, 1110 W. Green St., Urbana IL 61801-3080, U.S.A.}
\date{\today}

\begin{abstract}
We study the ${\mathbb Z}_{2}$ topologically ordered surface state of three-dimensional bosonic SPT phases with the discrete symmetries $G_{1} \times G_{2}$. It has been argued that the topologically ordered state cannot be realized on a purely two-dimensional lattice model. We carefully examine the statement and show that the suface state should break $G_{2}$ if the symmetry $G_{1}$ is gauged on the surface. This manifests the conflict of the symmetry $G_{1}$ and $G_{2}$ on the surface of the three-dimensional SPT phase. Given that there is no such phenomena in the purely two-dimensional model, it signals that the symmetries are encoded anomalously on the surface of the three-dimensional SPT phases and that the surface state can never be realized on the purely two-dimensional models. 
\end{abstract}

%\pacs{71.35.-y, 71.10.Pm, 73.20.-r}

\maketitle

According to Landau's theory of symmetry and symmetry breaking, 
there are two classes of the phases of matter: ordered phase and disordered phase. 
It has been suggested that there are at least three distinct types of the disordered phases. 
The first and most common phase is the classically disordered phase at a finite temperature. 
At the lowest temperature, there are two possible classes. In one possibility, the low-energy excitations are fraction of the fundamental particles, {\it e.g.,} electron, and are deconfined~\cite{Wen2004book}. 
For example, fractional quantum Hall states~\cite{Wen2004book, wenrev1, Wen1995, Wen95, Zhang1989, Lopez1991, Jain1989, Haldane1983} and spin liquids~\cite{Balents2010, ANDERSON1987, Wen2002, Lee2008a} belong to this class. 
The other quantum disordered states, which has been studied extensively recently, are so-called a Symmetry Protected Topological (SPT) phase~\cite{wenspt, Vishwanath2013, wenliu, Lu2012a, Oon2012, Lu2012b, Xu2012, Wang2013, Wang2013b, Ryu2012, Sule2013,classification2, Kitaev2009}. 
SPT phases do not support any fractional excitation and yet has no adiabatic path to the trivial atomic insulator,
once a set of symmetry conditions is strictly enforced. 
Electronic topological band insulators and topological superconductors~\cite{KaneRev, QiRev, Moore2007, Fu2007, Kane2005, qilong, essinmoore,classification2, Kitaev2009, Fu2008} examplify the SPT phases. 
Typically (but not necessarily) SPT phases have gapless spectrum at the boundary, 
and the gapless nature of the spectrum is protected by the symmetries. 

It has been shown that the gapless edge mode at the boundary of two-dimensional SPT phases can never be realized 
as a purely one-dimensional lattice model.
For free fermion systems, this is known 
in a form of a no-go theorem called  
fermion doubling problem.  
Even beyond non-interacting fermion systems,  
it has been shown that the gapless edge modes of SPT phases, 
with strict enforcement of the symmetry conditions,
suffer from various kinds of anomalies, signaling the impossibility of 
realizing them on an isolated one-dimensional lattice    
\cite{Ryu2012, Sule2013, Cappelli2013, Ringel2012b, Koch-Janusz2013} (see below for an example with $U(1)\times U(1)$ symmetry). 
For example, 
any quantum theory realized on an one-dimensional lattice model is expected to have the so-called modular invariance, 
which is an invariance under large coordinate transformations of 
the spacetime torus. 
It has been demonstrated that the gapless edge states of two-dimensional SPT phases 
violate the modular invariance once symmetry conditions are strictly enforced. 
\cite{Ryu2012, Sule2013}
%The modular non-invariance guarantees the stability of the edge state when it is coupled with a purely one-dimensional system. 

The surface of the three-dimensional SPT phases has more options than being gapless~\cite{Vishwanath2013, xuludwig, xu3dspt}. 
For example, 
it has been shown that the surface state of the three-dimensional SPT phase can be ${\mathbb Z}_{2}$ 
topologically ordered and gapped while keeping the symmetries intact~\cite{Vishwanath2013, xuludwig, xu3dspt}. 
In the ${\mathbb Z}_{2}$ ordered state, 
there are two types of excitations, so-called $e$-particle and $m$-particle, 
which have the mutually semionic statistics and transform projectively under the symmetries in the way that it cannot be realized on a purely two-dimensional lattice system. 

In this paper, we will examine the statement carefully and illustrate explicitly how the symmetries 
are encoded anomalously on the excitations of the surface of three-dimensional bosonic SPT phases. 
We will primarily be interested in the states with 
the on-site unitary discrete symmetries 
$G_{1} \times G_{2}$. We show that we cannot write a classical Chern-Simons theory of the surface consistent with the symmetries~\cite{Vishwanath2013, Wang2013}. Given this observation, one can ask if we need to give up all the information from the Chern-Simons theory. In fact, we can show that the topological $S$- and $T$-matrices defined by the Chern-Simons theory are invariant under the symmetries and thus are well-defined. Then one of our main findings is that once $G_{1}$ is gauged, 
$G_{2}$ is necessarily broken in that the topological $S$- and $T$- matrices, defined by another Chern-Simons theory for the {\it gauged} surface state, are not invariant under the symmetry $G_{2}$. 
In other words, the two symmetries $G_1$ and $G_2$, which are seemingly compatible with each other,
are actually in conflict  
on the surface of the three-dimensional SPT phase;
once gauged by $G_1$, the surface system is forced to be $G_1$ invariant,
and inevitably breaks $G_2$. More precisely, we pretend that the surface of the SPT phase {\it is} a purely two-dimensional state and will gauge $G_1$ to find $G_2$ broken.   
This should be contrasted with the purely two-dimensional models with the symmetries $G'_{1} \times G'_{2}$,
for which we show that there is no such phenomenon. 
Hence the symmetries are realized anomalously on the surface of the three-dimensional SPT phases.

Such conflict of the symmetries 
is also observed in the two-dimensional SPT phase. 
For concreteness, let us consider the edge of the quantum spin Hall phase with the symmetries 
$U_{em}(1) \times U_{S_{z}}(1)$
({\it i.e.,} instead of imposing time-reversal symmetry,
we impose condition, 
conservation of $z$-component of SU(2) spin on
quantum spin Hall insulators).  
There is a pair of counter-propagating fermion modes at the edge,
which is described schematically by the Lagrangian 
\beq
L = \psi^{\dagger}_{\uparrow} (i\partial_{t} - v\partial_{x})\psi_{\uparrow} + \psi^{\dagger}_{\downarrow} (i\partial_{t} + v\partial_{x})\psi_{\downarrow}.
\eeq 
The edge theory is the {\it half} of a conventional Luttinger liquid. As a way to diagnose the edge theory,  
we can imagine ``gauging'' $U_{em}(1)$ symmetry in the edge theory
\cite{Ryu2012, Levin2012, Sule2013};
this can be implemented by introducing twisting boundary conditions 
and then average over twisting angles. 
Only the states that are singlet under $U_{em}(1)$ survive the gauging.
In other words, gauging $U_{em}(1)$ is amount to projecting the Hilbert space into the sector 
singlet under $U_{em}(1)$ (for the cases where such gauging involves a discrete group,   
this procedure is also called ``orbifolding'' in the literature~\cite{CFTbook, PolchinskiStringbook}). 

To appreciate the effect of this gauging procedure, 
it is helpful to introduce a conformal field theory 
description~\cite{CFTbook, PolchinskiStringbook} 
at the edge of the quantum spin Hall phase 
which is parametrized by $x \in (0, L)$ and $x \sim x+ L$.
The left- and right-moving electron operators 
$\psi_{L,R}$ at the edge are given,
in terms of the chiral bosons $\phi_{L,R}$ as
\beq
\psi_{L} \sim e^{i \phi_{L}}, \quad \psi_{R} \sim e^{-i\phi_{R}}. 
\label{fermion}
\eeq
%On gauging the $U_{em}(1)$ symmetry or 
With enforcing the $U_{em}(1)$ symmetry, 
a set of ground states preserving $U_{em}(1)$ symmetry 
can be explicitly constructed using the state-operator correspondence. 
\beq
|GS \rangle_{\alpha} = \lim_{t\rightarrow -\infty} \exp \left(i \alpha(\phi_{L}(x,t) + \phi_{R}(x,t)) \right) |0 \rangle,  
\eeq
in which we have introduced a continuous parameter $\alpha \in {\mathbb R}$. 
The ground state imposes a boundary condition to the fermions \eqref{fermion} 
%which can be derived from the operator product expansion for $\sim \exp(i \nu \phi_{R/L})$~
\cite{CFTbook, PolchinskiStringbook}, 
\beq
\psi_{L/R}(0) = e^{2\pi i\alpha}\psi_{L/R}(L).  
\label{BC}
\eeq
For the ground state, one can ask the ``charge'' $S_{z}$ under $U_{S_{z}}(1)$ and the charge $Q$ under $U_{em}(1)$~\cite{CFTbook, Ryu2012}. 
\begin{align}
Q |GS \rangle_{\alpha} &= (\alpha -\alpha) |GS \rangle_{\alpha} =0,
\nonumber\\ 
S_{z}|GS \rangle_{\alpha} & = 
(\alpha + \alpha) |GS \rangle_{\alpha}  = 2\alpha |GS \rangle_{\alpha}.  
\end{align}
From the boundary conditions \eqref{BC}, we further see that 
a shift $\alpha \rightarrow \alpha +1$ is the symmetry of the system (which corresponds to a large gauge transformation of $U_{em}(1)$). 
We notice that this shift $\alpha \rightarrow \alpha+1$ leaves the total charge $Q$ of the ground state invariant (as it should be!). However, it does not leave the ``charge'' $S_{z}$ invariant, {\it i.e.}, after gauging $U_{em}(1)$,
the $U_{S_z}(1)$ symmetry is violated by quantum effects, 
and hence ``anomalous''. 
Alternatively, one can try to project the states with the symmetry $U_{S_{z}}(1)$ and then find, in this case,
that the symmetry $U_{em}(1)$ is broken (``anomalous''). 
So there is a conflict between the two symmetries $U_{em}(1) \times U_{S_{z}}(1)$. 
We further note that there is no such conflict in a conventional Luttinger liquid with the same symmetry $U_{em}(1) \times U_{S_{z}}(1)$ 
because each chirality comes with the two spin species and there is no net spin Hall effect. 
Thus the conflict of the symmetries implies that the symmetries are anomalously encoded at the edge of the SPT phases and 
that the edge theory can never be realized on a purely one-dimensional lattice. 

Motivated by these observations for SPT phases in (2+1) bulk dimensions, 
in this paper,  
we will study what we could learn by gauging symmetries
in SPT phases in (3+1) bulk dimensions, 
for the cases where their surface states are topologically ordered.  
We will show,
on the surface of three-dimensional SPT phases 
with the symmetries $G_{1} \times G_{2}$,  
there is a phenomenon similar to
``conflict of the symmetries'' illustrated above.  
As in (2+1)D SPT phases, 
we interpret the conflict as a sign that the surface state cannot emerge from a purely two-dimensional lattice model.  

A main difference from the case of (2+1)-dimensional SPT phases,
however, is that, since the surface is topologically ordered,
one of the symmetries, $G_1$, is a ``quantum'' (anyonic) symmetry, {\it i.e.,} the symmetry that cannot fully be incorporated at the level
of classical actions. Under $G_1$, quasi-particles (anyons) get transformed non-trivially,
while their topological properties (such as their braiding properties) are left invariant. 
On the other hand, the other symmetry $G_2$ is ``conventional'', 
which does not change the quasi-particle (anyon) types. 
We will show that once the conventional symmetry $G_2$ is gauged,
the resulting gauged theory is not invariant under the symmetry $G_1$ anymore.  

The rest of the paper is organized as following. In section I, we review the theory for the surface of a three-dimensional bosonic SPT phase. The $\mathbb{Z}_2$ topological order and the symmetries of the surface theory can be studied by a $O(4)$ non-linear sigma model that effectively reduces to a four component abelian Chern-Simons theory. We consider the bipartite ``conventional" and ``quantum" symmetry $G_1\times G_2$ that leaves the exchange and braiding information unchanged. As a comparison to non-holographic $(2+1)$D theories, we show the absence of gauging obstruction in section IB for a pure two-dimensional topological states with the same $G_1\times G_2$ symmetric structure. In section II, we study three prototypes of bosonic SPT with with (A) bipartite local unitary symmetries $\mathbb{Z}_2^A\times\mathbb{Z}_2^B$, (B) local unitary and time reversal symmetries $\mathbb{Z}_2\times\mathbb{Z}_2^T$, and (C) a tri-partite $\mathbb{Z}_2^A\times(\mathbb{Z}_2^B\times\mathbb{Z}_2^C)$ symmetry structure. We show that the gauging of the first ``conventional" $\mathbb{Z}_2$-symmetry would 
necessarily violate the remaining ``quantum" symmetries. This shows the ``quantum" anyonic symmetry is a topological obstruction to gauging the ``conventional" conterpart.

\section{Theory of Surface state}

It has been shown that the three-dimensional bosonic SPT phases can be successfully studied by the semi-classical O(5) non-linear sigma model supplemented by the topological $\Theta$-term~\cite{Bi2013, xu3dspt}. 
\beq
L = \frac{1}{2g^{2}} (\partial_{\mu} {\hat n})^{2} 
+ i\frac{2\pi}{64\pi^{2}} \varepsilon^{\mu\nu\lambda\rho}\varepsilon_{abcde} 
\hat{n}^{a}\partial_{\mu}
\hat{n}^{b}\partial_{\nu}
\hat{n}^{c}\partial_{\lambda}
\hat{n}^{d}\partial_{\rho}
\hat{n}^{e}
\label{NLSM}
\eeq
Different SPT phases correspond to the different ways of encoding the symmetries to the five-component unit vector ${\hat n}$~\cite{Bi2013, xu3dspt}. Then it can be shown that the surface is described by O(4) non-linear sigma model with $\Theta$-term at $\Theta = \pi$. Presumably there is a fixed point in which the low-energy physics is dominated solely by the topological $\Theta$-term. 
\beq
L = i\frac{\pi}{12\pi^{2}}\varepsilon^{\mu\nu\lambda}\varepsilon_{abcd} 
\hat{n}^{a}\partial_{\mu}
\hat{n}^{b}\partial_{\nu}
\hat{n}^{c}\partial_{\lambda}
\hat{n}^{d}
\eeq
The $\Theta$-term encodes the mutually semionic statistics of the fractionalized low-energy excitations $z_{e,a}, a=1,2$ and $z_{m,b}, b =1,2$ such that, 
\begin{align}
  &z^{\dagger}_{e} {\vec \sigma} z_{e} = (\hat{n}_{1}, \hat{n}_{2}, \hat{n}_{5}), \nonumber\\
  &z^{\dagger}_{m} {\vec \sigma} z_{m} = (\hat{n}_{3}, \hat{n}_{4}, \hat{n}_{5}).
\label{cp1representation}
\end{align}
The ${\mathbb Z}_{2}$ topologically ordered state on the surface can be obtained by condensation of the pair fields $\sim z^{*}z^{*} + h.c.$ of $z_{e}$'s or $z_{m}$'s~\cite{Bi2013, xu3dspt, Vishwanath2013}. We will assume that the mutually semionic statistics of $z_{e}$ and $z_{m}$ survives the condensation, and thus $z_{e}$ ($z_{m}$) becomes $e$-particle ($m$-particle) in the $Z_{2}$ topologically ordered state. Notice that $z_{e}$ and $z_{m}$ carry the symmetry index and transform as a doublet under the symmetries~\cite{Bi2013, xu3dspt, Vishwanath2013}. The symmetry action can be deduced from the relationships between $z$-fields and ${\hat n}$-fields and also from the solutions of the vortex in
 $\sim \hat{n}_{a} + i \hat{n}_{a+1}, a=1, 3$.  

Armed with this construction, we can proceed to construct an effective topological field theory for the topologically ordered state. 
Since the surface topological order is Abelian,
a natural candidate is an Abelian Chern-Simons theory. 
Due to the symmetry index carried by the $z$-fields, it is natural to consider $4$-component Chern-Simons theory. 
\beq
L = \frac{1}{4\pi} \varepsilon^{\mu\nu\lambda} {\vec a}^{T}_{\mu} K \partial_{\nu} {\vec a}_{\lambda} - {\vec a}^{T}_{\mu}{\vec J}_{\mu}
\label{ChernSimonsTheory}
\eeq
Here we used a vector notation for the gauge fields and the source currents, {\it i.e.,} ${\vec a}_{\mu}^{T} = (a^{1}_{\mu}, a^{2}_{\mu}, a^{3}_{\mu}, a^{4}_{\mu})$ and ${\vec J}_{\mu}^{T} = (J^{1}_{\mu}, J^{2}_{\mu}, J^{3}_{\mu}, J^{4}_{\mu})$, and $4 \times 4$ K-matrix. 
\beq
K = 
\left[ 
\begin{array}{cccc} 
0 & 0 & 1 & 1\\
0 & 0 & 1 &-1\\
1 & 1 & 0 & 0\\
1& -1 & 0 &0  
\end{array} 
\right]
\label{Kmatrix}
\eeq
The $K$-matrix correctly reproduces the statistics between the excitations and the topological degeneracies. 
The relationship between the current $J^{a}_{\mu}$ and the excitations can be explicitly written as 
\begin{align}
&J^{1}_{\mu} \sim -i(z^{\dagger}_{e,1}\partial_{\mu}z_{e,1} - h.c.) \nonumber\\
&J^{2}_{\mu} \sim -i(z^{\dagger}_{e,2}\partial_{\mu}z_{e,2} - h.c.) \nonumber\\
&J^{3}_{\mu} \sim -i(z^{\dagger}_{m,1}\partial_{\mu}z_{m,1} - h.c.) \nonumber\\
&J^{4}_{\mu} \sim -i(z^{\dagger}_{m,2}\partial_{\mu}z_{m,2} - h.c.) 
\label{Current}
\end{align}

\subsection{Symmetries of surface state}
Central to our discussion below is the way global symmetries are encoded in
the effective field theory for the surface.  
In ordinary situations
in 
quantum field theories, 
and many-body quantum systems defined on a lattice,  
symmetries are discussed at the level of classical actions or Lagrangians.
One may ask, subsequently, if the symmetries are preserved or not 
at the level of quantum mechanics,
{\it i.e.}, if the symmetries suffer from an anomaly or not.
For surfaces of SPT phases, 
however, 
as demonstrated through various examples~\cite{Vishwanath2013, Wang2013}, 
it may not be possible to write down a classical theory (action) for the surface respecting the symmetries,
at least in the naive way. 
In the following, therefore, 
we will distinguish {\it classical} and {\it quantum} symmetries.

%Then 
The symmetry action on the currents $J^{a}$ ($a = 1 \cdots 4$) can be deduced from the symmetry action on $z$-fields, ${\vec J}_{\mu} \rightarrow X {\vec J}_{\mu}$, $X \in GL(4, {\mathbb Z})$. Under the symmetry, the effective theory \eqref{ChernSimonsTheory} transforms into, 
\begin{align}
X: L &\rightarrow  \frac{1}{4\pi} \varepsilon^{\mu\nu\lambda} {\vec a}^{T}_{\mu} K \partial_{\nu} {\vec a}_{\lambda} - {\vec a}^{T}_{\mu}X{\vec J}_{\mu} 
\nonumber\\
&= \frac{1}{4\pi} \varepsilon^{\mu\nu\lambda} {\vec b}^{T}_{\mu} {\tilde K} \partial_{\nu} {\vec b}_{\lambda} - {\vec b}^{T}_{\mu}{\vec J}_{\mu},
\end{align}
in which we changed the variable via ${\vec b}_{\mu} = X^{T}{\vec a}_{\mu}$ or ${\vec a}_{\mu} = (X^{T})^{-1}{\vec b}_{\mu}$ and ${\tilde K} = X^{-1}K (X^{-1})^{T}$.

The Chern-Simons theory 
in the absence of the sources $J_{\mu}$
is classically invariant when $\tilde{K}=K$. 
We say the Chern-Simons theory has a classical symmetry $X$.

On the other hand, 
notice that $X \in GL(4, {\mathbb Z})$ and this implies that the Chern-Simons theories with $K$ and ${\tilde K} =  X^{-1}K (X^{-1})^{T}$ define the same topological order. 
Thus,
while the classical action \eqref{ChernSimonsTheory} might not be invariant 
under the symmetry transformation, 
the quantum information encoded by the Chern-Simons theory \eqref{ChernSimonsTheory} 
may still be invariant under the symmetry action. 
More precisely,
we consider topological $S$- and $T$- matrices of the topological state and show that the matrices are invariant under the symmetry transformations. 
We call $X$ a quantum (anyonic) symmetry
if $K \neq  X^{-1}K (X^{-1})^{T}$ 
while topological $S$- and $T$- matrices are invariant under $X$ (there is a closely related but different definition for the anyonic symmetry in the reference~\cite{KhanTeoHughes}), {\it i.e.} $X^{T} K^{-1} X = K ^{-1}$ mod $N$ where $N$ is a matrix with the integers ${\mathbb Z}$ for the off-diagonal elements and $2{\mathbb Z}$ for the diagonal elements.

We now explain the $S$- and $T$- matrices which will play an important role in our discussion. 
Quasiparticle excitations of the topological field theory \eqref{ChernSimonsTheory} are labeled as integer lattice vectors $\vec{u}$ in $\Gamma^\ast=\mathbb{Z}^N$. Each has the statistical angle $\theta_{\vec{u}}=\pi\vec{u}^TK^{-1}\vec{u}$, 
which corresponds to the phase of exchange between a pair of identical quasiparticles and the phase of $2\pi$-twist of a single one.
The braiding phase between quasiparticles $\vec{u}$ and $\vec{v}$ is given by the pairing $2\pi\vec{u}^TK^{-1}\vec{v}$. 
There is a subset of quasiparticles that are local with respect to all quasiparticles. 
They live in the sublattice $\Gamma=K\mathbb{Z}^N$ under the image of the $K$-matrix so that their braiding phases with the rest are integer multiples of $2\pi$.
We consider bosonic theories where local particles are bosons ($\theta\equiv0$ mod $2\pi\mathbb{Z}$) with even diagonal 
$K$-matrix entries. Anyons are equivalent up to local particles and are labeled by equivalent classes of lattice vectors $[\vec{u}]=\vec{u}+\Gamma$. They live in the finite abelian quotient group $\mathcal{A}=\Gamma^\ast/\Gamma$ so that the anyonic fusion structure is identical to the group multiplication. The braiding and exchange information is encoded by the two unitary matrices. 
\begin{align}
S_{[\vec{u}][\vec{v}]} =\frac{1}{\mathcal{D}}e^{2\pi i\vec{u}^TK^{-1}\vec{v}}, \quad T_{[\vec{u}][\vec{v}]} &=\delta_{[\vec{u}][\vec{v}]}e^{i\theta_{\vec{u}}}
\label{TopMat}
\end{align} 
The matrices projectively represent the modular group $SL(2;\mathbb{Z})$, where the total quantum dimension is related to the number of anyon types by $\mathcal{D}^2=|\det(K)|=|\mathcal{A}|$ in the abelian theory. We can compactly define $S$- and $T$- matrices for a given $K$-matrix. 
\beq
S = \frac{1}{\mathcal{D}} \exp(2i\pi K^{-1}), \quad T = \text{diag}[ \exp(i\pi K^{-1}) ]
\eeq

We emphasise again that a quantum symmetry $X$ is not a symmetry of the $K$-matrix.
For the surface physics of 3D SPTs, 
this has been thought of as the signal that the topological phase 
with $X$ can never be realized on the pure two dimensional lattice 
systems because one cannot write down a Chern-Simons theory consistent with the symmetry operation.

We would like to elaborate this point further in this paper. In fact, the invariance of the $S$- and $T$- matrices \eqref{TopMat} under the symmetries hints us that we can write down the symmetric partition function which can generate the $S$- and $T$- matrices for the surface of the SPT phase. Instead of the conventional partition function $Z^{CS}_{K}$ for the Chern-Simons theory with the fixed $K$ matrix, 
\begin{align}
Z^{CS}_{K}[J^{a}_{\mu}] &=  \int \prod_{a = 1 \cdots 4} D a^{a}_{\mu} \exp(i \int dt d^{2}x L^{CS}_{K}[J^{a}_{\mu}]), \nonumber\\
L^{CS}_{K}[J^{a}_{\mu}] &= \frac{1}{4\pi} \varepsilon^{\mu\nu\lambda} {\vec a}^{T}_{\mu}K \partial_{\nu} {\vec a}_{\lambda} - {\vec a}^{T}_{\mu}{\vec J}_{\mu}, 
\end{align}
let us consider a following partition function which has a sum over $Y \in G \subseteq GL(4; {\mathbb Z})$, the symmetry group of the SPT phase, to describe the surface of the three-dimensional SPT.  
\begin{align}
Z^{SPT}[J^{a}_{\mu}] &= \sum_{Y} \int \prod_{a = 1 \cdots 4} D a^{a}_{\mu} \exp(i \int dt d^{2}x L[Y; J^{a}_{\mu}]) 
\nonumber\\
L[Y; J^{a}_{\mu}] &= \frac{1}{4\pi} \varepsilon^{\mu\nu\lambda} {\vec a}^{T}_{\mu} Y^{T}KY \partial_{\nu} {\vec a}_{\lambda} - {\vec a}^{T}_{\mu}{\vec J}_{\mu}, Y \in GL(4, {\mathbb Z})
\label{partitionfunction}
\end{align}
In the case that there is no continuous symmetry and no edge state, the only physical observables are the phases when two quasi-particles are braided. For the concreteness, let us consider the geometry $S^{3}$~\cite{Witten1989}. The phases can be computed by considering the source currents $J^{a}_{\mu} ({\vec x}, t)$ and $J^{b}_{\mu}({\vec x}, t)$ with the linking number $L$. In a conventional Chern-Simons theory, the (self)-linking number of the multi-component $U(1)$-current $J^a_\mu(x,t)$ is given by the integral. 
\begin{align}
L[J^a_\mu]=\int_{S^3}
d^2x dt\,
\varepsilon^{\mu\nu\lambda} (K^{-1})_{ab}J^a_\mu\partial_\nu J^b_\lambda
\end{align} 
This identifies with the braiding phase $\frac{Z[J^a_\mu]}{Z[0]}=e^{2\pi iL[J^a_\mu]}$. In terms of the partition function \eqref{partitionfunction}, we can also compute the statistical phases.  
\begin{align}
S_{ab} &= \frac{Z^{SPT}[J^{a}_{\mu}]}{Z^{SPT}[0]}, \nonumber\\
&= \frac{\int D[Y] Z^{CS}_{Y^{T}KY}[J^{a}_{\mu}] }{\int D[Y] Z^{CS}_{Y^{T}KY}[0]} = \frac{\int D[Y] Z^{CS}_{Y^{T}KY}[0] e^{2\pi i(K^{-1})^{ab} L} }{\int D[Y] Z^{CS}_{Y^{T}KY}[0]},\nonumber\\
&= \frac{\int D[Y] Z^{CS}_{K}[0] e^{2\pi i(K^{-1})^{ab} L}}{\int D[Y] Z^{CS}_{K}[0]} = \exp(2\pi i(K^{-1})^{ab} L), 
\end{align}
where we have used the fact that $Z^{CS}_{Y^{T}KY}[0] = Z^{CS}_{K}[0]$ on the geometry $S^{3}$ or $S^{1} \times M$~\cite{Witten1989} (also notice that the $S$-matrix obtained above is consistent with \eqref{TopMat}). Furthermore, it is not difficult to see that the partition function $Z^{SPT}[J^{a}_{\mu}]$ and its observables $S$- and $T$-matrices are invariant under the quantum symmetry (the $S$- and $T$- matrices correspond to the braiding of two excitations in the topological phase). With the sum over $Y \in G \subseteq GL(4, {\mathbb Z})$, we can write down a partition function consistent with the symmetries though we cannot write a classical Chern-Simons theory. From the view point of the partition function and the S- and T- matrices \eqref{TopMat}, it is unclear in which respect the surface of the three-dimensional SPT phase is anomalous or how the surface of the three-dimensional SPT phase is different from the purely two-dimensional models.  

We will show that gauging a part of the symmetries of the SPT phase~\cite{Lu2013} with the symmetry $Y \times X$ can help to resolve the question. 
Usually there is a preferred symmetry, {\it e.g.,} $Y$, to be gauged on the surface of a three-dimensional SPT phase. 
We will gauge the symmetry which is a classical symmetry, {\it i.e.,} the symmetry of the K-matrix.

The gauging procedure can be thought of as deconfining a new excitation which is a fraction of the original excitation. 
Because we would like to have an integral anyon lattice, we need to find a new $K$-matrix and identify the symmetry actions 
in the new anyon lattice. 
This allows us to study the symmetry operations on the new $S$- and $T$- matrices based on the new $K$-matrix. If the $S$- and $T$- matrices were found invariant under the symmetry $X$, then we could have come up with the symmetric partition function \eqref{partitionfunction} which can be used to generate the $S$- and $T$- matrices for the {\it gauged} theory as the above. However we will show that the symmetry $X$ of the SPT phase is no longer a (quantum) symmetry of the {\it gauged} theory. In fact, the symmetry $X$ does not leave the anyon lattice invariant and can never be the symmetry of the $S$- and $T$- matrices. 

\subsection{Comparison to (2+1)-dimensional topological phases}\label{twoD}

This should be compared with the two-dimensional models. 
Let us imagine a topologically ordered state realized on the purely two-dimensional lattice models with the symmetries $G_{1} \times G_{2}$. 
We begin with the $\mathbb{Z}_2$ topological phase (a $\mathbb{Z}_2$ discrete gauge theory) described by a $K$-matrix with global symmetry $G_1\times G_2$ in which $G_{1}$ is ``conventional'' in the sense of the reference~\cite{Lu2013}, 
{\it i.e.,} $G_{1}$ does not change the excitations type and induces only the phase rotations. 
To manifest the action of the symmetries, 
we consider the chiral boson fields $\vec \phi$ living at the edge of the topologically ordered surface state. The symmetries are projectively encoded on the edge excitations as
\begin{align}
G_{1}&: {\vec \phi} \rightarrow {\vec \phi} + \delta {\vec \phi},  \nonumber\\ 
G_{2}&: {\vec \phi} \rightarrow X{\vec \phi} + \delta {\vec \theta}. 
\label{symmetry_intro}
\end{align}
The $X$ should be an element of $GL(N, {\mathbb Z})$ and should be compatible with the $K$-matrix~\cite{Lu2013, Lu2012a}, $X^{T}KX = K$. We notice that $X \delta {\vec \phi} = \delta {\vec \phi}$ by requiring $G_{1}$ and $G_{2}$ commute. Next we ``promote" the conventional symmetry $G_1$ into a local symmetry by extending the gauge group from $\mathbb{Z}_2\hookrightarrow\mathbb{Z}_2\times G_1$. The gauging procedure extends the topological field theory by introducing new $G_1$ fluxes. We are interested in minimal extensions so that no extra $G_1$ charges other than the original quasiparticles are added to the theory. For simplicity, we assume $G_1$ is finite abelian and, without loss of generality, generated by a single element, i.e. $G_1=\mathbb{Z}_k$ \eqref{symmetry_intro}. As $G_1$ is of finite order, $k$ fluxes would fuse to the trivial flux, and corresponds to a quasiparticle in the ungauged topological phase. We represents the additional $G_1$ flux by the fractional lattice vector $\vec{l}_v$ so that $k\vec{l}_v$ lives in $\Gamma=\mathbb{Z}^N$. According to \eqref{symmetry_intro}, the braiding phase of the quasiparticle $\psi_{\vec{u}}\sim e^{i\vec{u}^T\vec{\phi}}$ with the $G_1$ flux is given by $e^{2\pi i\vec{u}^TK^{-1}\vec{l}_v}=e^{i\vec{u}^T\delta\vec\phi}$ which has solutions, 
\begin{align}
\vec{l}_v=\frac{1}{2\pi}K\delta\vec\phi,
\end{align} 
up to integral vectors in $\mathbb{Z}^N$ (for more complete discussion on gauging symmetries in a Chern-Simons theory, we refer the reader to the references~\cite{Lu2013, Lu2012a}). Now we ask if $G_{2}$ leaves the anyon lattice invariant when $G_{1}$ is gauged. Notice that $G_{2}$ will leave the integral lattice invariant and thus it is enough to see how ${\vec l}_{v} = \frac{K}{2\pi} \delta {\vec \phi}$ transform under $G_{2}$. 
\begin{align}
G_{2}: {\vec l}_{v} = \frac{K}{2\pi} \delta {\vec \phi} &\rightarrow X^{T}\frac{K}{2\pi} \delta {\vec \phi}, \nonumber\\
&= X^{T}\frac{K}{2\pi}X X^{-1}\delta {\vec \phi}, \nonumber\\
&= \frac{K}{2\pi} \delta {\vec \phi} = {\vec l}_{v}.
\end{align}
Thus the anyon lattice in the gauged theory is invariant under the symmetry $G_{2}$ in the purely two-dimensional lattice models.

We also consider the symmetry group $G$ generated by $G_{1}$ and $G_{2}$ which do not commute. As before we assume $G_{1}$ is ``conventional''~\cite{Lu2013} and shifts the phases of the quasi-particle excitations.  
\begin{align}
G_{1}&:\vec{\phi} \rightarrow {\vec \phi} + \delta{\vec \phi} \nonumber\\
G_{2}&: \vec{\phi} \rightarrow X {\vec \phi} +\delta \vec{\theta}
\end{align}
We require $G_{2}$ to be the symmetry of the $K$-matrix, {\it i.e.,} $X \in GL(N, {\mathbb Z})$, and $X^{T}K = KX^{-1}$~\cite{Lu2013, Lu2012a}. Because $G_{1}$ and $G_{2}$ do not commute, $X \delta \vec{\phi} \neq \delta \vec{\phi}$. Thus we need to be more careful in gauging the conventional symmetry group ``generated'' by $G_{1}$ and first find a normal subgroup $H$ generated by $\{ G^{m}_{2}G^{n}_{1}G^{-m}_{2}, (n,m) \in {\mathbb Z}^{2} \}$ (notice that $G_{1} \in H$). 
\beq
G^{m}_{2}G^{n}_{1}G^{-m}_{2}: \phi \rightarrow \phi + n X^{m}\delta {\vec \phi}
\eeq
Thus there are many different generators $\{ X^{n}\delta \vec{\phi}, n \in {\mathbb Z} \}$ of the phase shifts which we need to gauge. Thus it is natural to deconfine all the different anyonic excitations corresponding to $\{ X^{n}\delta \vec{\phi}, n \in {\mathbb Z} \}$ when we gauge a normal subgroup symmetry $H$ in $G$.  
\beq
\vec{l}_{n, v} = \frac{K}{2\pi}X^{n}\delta \vec{\phi}
\eeq
We ask if the anyon lattice of this gauged theory is invariant under $G_{2}$. As the integral lattice is invariant under $X$, we only need to check if the quasi-particle lattice generated by $\{ {\vec l}_{n,v}, n \in {\mathbb Z} \}$ is invariant under $G_{2}$. It is not difficult to check that the quasi-particle lattice is invariant under the symmetry ${\mathbb Z}_{2}$ 
since $X^{T} {\vec l}_{n,v} = {\vec l}_{n-1,v}$ because of $X^{T}K = KX^{-1}$.  

Thus the anyon lattice in the gauged theory is invariant under the symmetry $G_{2}$ in the both cases. Furthermore, $X^{T}KX =K$ with the invariance of the anyon lattice guarantees the invariance of the $S$- and $T$- matrices \eqref{TopMat}~\cite{KhanTeoHughes}. We will show that this is not the case on the surface of the three-dimensional SPT phases. 

\section{Examples}
We carefully go through one example with the symmetry ${\mathbb Z}^{A}_{2} \times {\mathbb Z}^{B}_{2}$~\cite{Bi2013} and illustrate the strategies. For other symmetric cases ${\mathbb Z}_{2}\times {\mathbb Z}^{T}_{2}$ and ${\mathbb Z}^{A}_{2} \times {\mathbb Z}^{B}_{2} \times {\mathbb Z}^{C}_{2}$~\cite{Bi2013}, we will show the results only.   

\subsection{${\mathbb Z}^{A}_{2}\times {\mathbb Z}^{B}_{2}$ symmetric case}
On the surface of the three-dimensional SPT phase with the ${\mathbb Z}^{A}_{2}\times {\mathbb Z}^{B}_{2}$ symmetries, the symmetries are projectively represented on the excitations~\cite{Bi2013}. 
\begin{align}
{\mathbb Z}^{A}_{2} &: z_{e} \rightarrow i\sigma^{z} z_{e}\nonumber\\ 
&: z_{m} \rightarrow z_{m} \nonumber\\
{\mathbb Z}^{B}_{2} &: z_{e} \rightarrow \sigma^{x} z_{e}\nonumber\\ 
&: z_{m} \rightarrow i\sigma^{y}z^{*}_{m}
\label{symmetry1}
\end{align} 
Notice that ${\mathbb Z}^{A}_{2}$ and ${\mathbb Z}^{B}_{2}$ 
do not commute when acting on the fractional excitations 
$z_{e}$ and $z_{m}$. 
(It should however be noted that 
the symmetries do commute~\cite{Bi2013} 
when acting on the $O(5)$ field ${\hat n}$ appearing in the non-linear sigma model \eqref{NLSM}). 
One might naively think that the fact that the two symmetries 
do not commute implies the conflict of the symmetries in the gauge theory. 
We have seen however that this is not the case as illustrated for the purely two-dimensional lattice models in \ref{twoD}.

Both the symmetries are unitary and on-site. The symmetries on the currents $J \sim iz^{*}\partial z + h.c.$ \eqref{Current} are represented by the $4 \times 4$ matrices. 
\begin{align}
{\mathbb Z}^{A}_{2} &: J^{a}_{\mu} \rightarrow \delta^{ab}J^{b}_{\mu}, \nonumber\\
{\mathbb Z}^{B}_{2} &: J^{a}_{\mu} \rightarrow X^{ab}J^{b}_{\mu}, 
\end{align}
where 
\beq
X = 
\left[ 
\begin{array}{cccc} 
0 & 1 & 0 & 0\\
1 & 0 & 0 & 0\\
0 & 0 & 0 &-1\\
0&  0 &-1 & 0  
\end{array} 
\right]. 
\label{op1}
\eeq
There are a few points worth commenting. 
First of all, one might worry that ${\mathbb Z}^{A}_{2}$ is represented trivially $\sim \delta^{ab}$ on the currents, and hence the symmetry ${\mathbb Z}^{A}_{2}$ is almost like ``doing nothing'' if we insist on working with the effective theory. 
However, ${\mathbb Z}^{A}_{2}$ is merely a global phase rotation (or global gauge transformation) and thus it is natural that the global phase rotation does not appear explicitly in the Chern-Simons theory (which is a gauge theory and the gauge theory is only sensitive to the gradient in the phases of the matter fields). To properly take ${\mathbb Z}^{A}_{2}$ symmetry into account in the gauge theory, we are led to consider gauging ${\mathbb Z}^{A}_{2}$ symmetry. Roughly speaking, gauging the symmetry can be understood as performing ${\mathbb Z}^{A}_{2}$ locally and thus generates the gradient in the phases which do appear in the gauge theory. 

Secondly it is trivial to see that ${\mathbb Z}^{A}_{2}$ is the symmetry of the $K$-matrix \eqref{Kmatrix} and thus ${\mathbb Z}^{A}_{2}$ does not change the $S$- and $T$- matrices because $\delta^{ab} \in GL(4, {\mathbb Z})$ and $K$ is invariant under the symmetry. 

On the other hand, ${\mathbb Z}^{B}_{2}$ is not a symmetry of $K$-matrix but only a quantum symmetry. This can be explicitly checked by computing ${\tilde K} = X^{-1}K (X^{-1})^{T}$. 
\beq
{\tilde K} = \left[ 
\begin{array}{cccc} 
0 & 0 & 1 & -1\\
0 & 0 & -1 &-1\\
1 & -1 & 0 & 0\\
-1& -1 & 0 &0  
\end{array} 
\right] 
\eeq 
It is not difficult to see that ${\tilde K}$ is different from $K$ but has the same topological order and topological $S$- and $T$- matrices as $K$. Thus we find that $S$- and $T$- matrices are invariant under both the symmetries ${\mathbb Z}^{A}_{2}\times {\mathbb Z}^{B}_{2}$. 

Now we proceed to gauge the ${\mathbb Z}^{A}_{2}$ symmetry. We closely follow the prescriptions previously introduced in ref ~\onlinecite{Lu2013}. First, we identify a new anyon excitation $l_{v}$ in the charge lattice. 
\beq
{\vec l}_{v} = K \frac{ \delta {\vec \phi}}{2\pi} = (0,0,0,\frac{1}{2})^{T},
\label{deconfined_vortex1}
\eeq
in which we have used $\delta {\vec \phi} = (\pi/2, -\pi/2, 0,0)^{T}$ from \eqref{symmetry1}. Notice that $X \delta {\vec \phi}$ is simply $-\delta {\vec \phi}$, and thus it is enough to deconfine ${\vec l}_{v}$ above \eqref{deconfined_vortex1} to gauge the ${\mathbb Z}^{A}_{2}$ symmetry. The new excitation ${\vec l}_{v}$ has the mutual statistics with the original anyon excitations to incorporate the ``gauged'' symmetry ${\mathbb Z}^{A}_{2}$. With this new excitation, the allowed anyon excitation in the gauged theory can be represented. 
\beq
{\vec l} = (n_{1},n_{2},n_{3}, n_{4}+\frac{n_{v}}{2})^{T}
\label{AnyonLattice1}
\eeq
with the statistical angle $\theta = \pi {\vec l}^{T} K^{-1} {\vec l} $. Notice that the ${\mathbb Z}^{B}_{2}$ symmetry operation $X$ \eqref{op1} does not leave the anyon lattice \eqref{AnyonLattice1} invariant because it maps $(0,0,0,1/2)^{T}$, the new anyonic excitation, to $(0,0,-1/2,0)^{T}$ which is not allowed in the theory. This already signals that the symmetry ${\mathbb Z}^{B}_{2}$ is not the symmetry of the gauged theory. 

To understand the gauged theory better, we extend the anyon lattice (which contains the `fractional' excitation ${\vec l}_{v} = (0,0,0,1/2)^{T}$) with $K$ to the integral anyon lattice with the new $K_{g}$ matrix to describe the theory properly. To rescale the lattice and the $K$-matrix properly, we need to find a matrix $M$ such that,
\beq
{\vec l} = (n_{1},n_{2},n_{3}, n_{4}+\frac{n_{v}}{2})^{T} = M {\vec m}, \quad {\vec m} \in {\mathbb Z}^{4}.
\eeq 
We choose $M$ in the way that all the excitations defined by $l$ and the excitations defined by ${\vec m}$ are equivalent ($M$ is bijective).  
\beq
M = \left[ 
\begin{array}{cccc} 
1 & 0 & 0 &0\\
0 & 1 & 0 &0\\
0& 0& 1 & 0\\
0& 0 & 0 &\frac{1}{2}  
\end{array} 
\right] 
\eeq
By requiring the statistical angle defined by ${\vec l}$ and ${\vec m}$ to be the same, we obtain the condition for the new $K_{g}$ matrix for the gauged theory. 
\beq
M^{T}K^{-1}M = K^{-1}_{g}
\eeq

As a result of gauging the symmetry ${\mathbb Z}^{A}_{2}$, we end up with the theory, 
\beq
K_{g} = \left[ 
\begin{array}{cccc} 
0 & 0 & 1 &1\\
0 & 0 & 2 &-2\\
1& 2& 0 & 0\\
1& -2 & 0 &0  
\end{array} 
\right],  
\eeq
with the integral anyonic lattice ${\vec m} \in {\mathbb Z}^{4}$. 

We now ask if this gauged $K_{g}$-matrix can generate well-defined $S$- and $T$- matrices invariant under ${\mathbb Z}^{B}_{2}$ as the case before gauging. In this gauged theory, the original excitations are represented by $z_{e1} \sim (1,0,0,0)^{T}$, $z_{e2} \sim (0,1,0,0)^{T}$, $z_{m1} \sim (0,0,1,0)^{T}$, and $z_{m2} \sim (0,0,0, 2)^{T}$. From this information, we can easily read off the action ${\tilde X}$ of the symmetry ${\mathbb Z}^{B}_{2}$ as before. 
\beq
{\tilde X} = \left[ 
\begin{array}{cccc} 
0 & 1 & 0 &0\\
1 & 0 & 0 &0\\
0& 0& 0 & -\frac{1}{2}\\
0& 0 & -2 &0  
\end{array} 
\right]
\eeq
We note that ${\tilde X}$ is not an element of $GL(4, {\mathbb Z})$ and does not leave the anyon lattice invariant. Furthermore, it is now trivial to see that ${\tilde S}$- and ${\tilde T}$- matrices, defined by $K_{g}$, are not invariant under ${\tilde X}$. This concludes that the symmetry ${\mathbb Z}^{B}_{2}$ is broken if we gauge the symmetry ${\mathbb Z}^{A}_{2}$. 

\subsection{${\mathbb Z}_{2}\times {\mathbb Z}^{T}_{2}$ symmetric case}
On the surface of the three-dimensional SPT phase with the ${\mathbb Z}_{2}\times {\mathbb Z}^{T}_{2}$ symmetries, the symmetries are represented as following~\cite{Bi2013}. 
\begin{align}
{\mathbb Z}_{2} &: z_{e} \rightarrow i\sigma^{z} z_{e}\nonumber\\ 
&: z_{m} \rightarrow i\sigma^{z} z_{m} \nonumber\\
{\mathbb Z}^{T}_{2} &: z_{e} \rightarrow i\sigma^{y} z_{e}\nonumber\\ 
&: z_{m} \rightarrow i\sigma^{y}z_{m}
\label{symmetry2}
\end{align} 
The ${\mathbb Z}_{2}$ symmetry is unitary and on-site, and ${\mathbb Z}^{T}_{2}$ is local and anti-unitary. The symmetries on the current $J \sim iz^{*}\partial z + h.c.$ \eqref{Current} are represented by the $4 \times 4$ matrices as before. 
\begin{align}
{\mathbb Z}_{2} &: J^{a}_{\mu} \rightarrow \delta^{ab}J^{b}_{\mu}, \nonumber\\
{\mathbb Z}^{T}_{2} &: J^{a}_{\mu} \rightarrow \eta^{\mu\nu} X^{ab}_{T}J^{b}_{\nu}, \nonumber\\ 
&:K \rightarrow -K
\end{align}
where $\eta^{\mu\nu} = 1$ if $\mu =\nu= 0 $ and $-1$ if $\mu =\nu= 1,2$
and
\beq
X_{T} = 
\left[ 
\begin{array}{cccc} 
0 & 1 & 0 & 0\\
1 & 0 & 0 & 0\\
0 & 0 & 0 &1\\
0&  0 &1 & 0  
\end{array} 
\right] \in GL(4; {\mathbb Z})
\label{op2}
\eeq
Notice that ${\mathbb Z}_{2}$ is the symmetry of the $K$-matrix \eqref{Kmatrix} and thus does not change the $S$- and $T$- matrices because $\delta^{ab} \in GL(4, {\mathbb Z})$ and $K$ is invariant under the symmetry. On the other hand, ${\mathbb Z}^{T}_{2}$ is not a symmetry of 
the $K$-matrix but only a quantum symmetry. 

Now we proceed to gauge the ${\mathbb Z}_{2}$ symmetry. As before, we begin with identifying a new anyon excitation $l_{v}$ in the charge lattice~\cite{Lu2013}. 
\beq
{\vec l}_{v} = K \frac{ \delta {\vec \phi}}{2\pi} = (0,\frac{1}{2},0,\frac{1}{2})^{T},
\label{deconfined_vortex2}
\eeq
in which we have used $\delta {\vec \phi} = (\pi/2, -\pi/2, \pi/2, -\pi/2)^{T}$ from \eqref{symmetry2}. Notice that $X_{T} \delta {\vec \phi}$ is simply $\delta {\vec \phi}$, and thus it is enough to deconfine ${\vec l}_{v}$ above \eqref{deconfined_vortex2} to gauge the ${\mathbb Z}_{2}$ symmetry. With this new excitation, the allowed anyon excitation in the gauged theory can be represented. 
\beq
{\vec l} = (n_{1},n_{2}+\frac{n_{v}}{2},n_{3}, n_{4}+\frac{n_{v}}{2})^{T}
\label{AnyonLattice2}
\eeq
with the statistical angle $\theta = \pi l^{T} K^{-1} l $. Notice that the ${\mathbb Z}^{T}_{2}$ symmetry operation $X$ \eqref{op2} does not leave the anyon lattice \eqref{AnyonLattice2} invariant because it maps $(0,1/2,0,1/2)^{T}$, the new anyonic excitation, to $(-1/2,0,-1/2,0)^{T}$ which is not allowed in the theory. This already signals that the symmetry ${\mathbb Z}^{T}_{2}$ is not the symmetry of the gauged theory. 

We extend the anyon lattice with $K$ to the integral anyon lattice with the new $K_{g}$ matrix to describe the theory properly. 
To rescale the lattice and the $K$-matrix, we find a matrix $M$ such that,
\beq
{\vec l} = (n_{1},n_{2}+\frac{n_{v}}{2},n_{3}, n_{4}+\frac{n_{v}}{2})^{T} = M {\vec m}, \quad {\vec m} \in {\mathbb Z}^{4}.
\eeq 
We choose $M$ which is bijective.  
\beq
M = \left[ 
\begin{array}{cccc} 
0 & 1 & 0 &0\\
\frac{1}{2} & 0 & 1 &0\\
0& 0& 0 & 1\\
\frac{1}{2}& 0 & 0 &0  
\end{array} 
\right] 
\eeq
From $M$, we obtain the new $K_{g}$ matrix for the gauged theory by requiring,  
\beq
M^{T}K^{-1}M = K^{-1}_{g}. 
\eeq

As a result of gauging the symmetry ${\mathbb Z}_{2}$, we end up with the theory, 
\beq
K_{g} = \left[ 
\begin{array}{cccc} 
0 & 2 & -2 &0\\
2 & 0 & -1 &1\\
-2& -1& 2 & 1\\
0& 1 & 1 &0  
\end{array} 
\right],  
\eeq
with the integral anyonic lattice ${\vec m} \in {\mathbb Z}^{4}$. 

We now ask if this gauged $K_{g}$-matrix can generate well-defined $S$- and $T$- matrices invariant under ${\mathbb Z}^{T}_{2}$ as the case before gauging. The original excitations are represented by $z_{e1} \sim (0,1,0,0)^{T}$, $z_{e2} \sim (0,0,1,0)^{T}$, $z_{m1} \sim (0,0,0,1)^{T}$, and $z_{m2} \sim (-2,0,-1, 0)^{T}$. From this information, we can find the action ${\tilde X}_{T}$ of the symmetry ${\mathbb Z}^{T}_{2}$ as before. 
\beq
{\tilde X}_{T} = \left[ 
\begin{array}{cccc} 
0 & 0 & 0 &2\\
\frac{1}{2} & 0 & 1 &0\\
0& 1& 0 & -1\\
\frac{1}{2}& 0 & 0 &0  
\end{array} 
\right]
\eeq
We note that ${\tilde X}_{T}$ is not an element of $GL(4, {\mathbb Z})$ and does not leave the anyon lattice invariant. We can also check that ${\tilde S}$- and ${\tilde T}$- matrices \eqref{TopMat} defined by $K_{g}$ are not invariant under the action of ${\tilde X}_{T}$. This concludes that the symmetry ${\mathbb Z}^{T}_{2}$ is broken if we gauge the symmetry ${\mathbb Z}_{2}$. 

\subsection{${\mathbb Z}^{A}_{2} \times {\mathbb Z}^{B}_{2} \times {\mathbb Z}^{C}_{2}$ symmetric case}
On the surface of the three dimensional SPT phase with the ${\mathbb Z}^{A}_{2} \times {\mathbb Z}^{B}_{2} \times {\mathbb Z}^{C}_{2}$ symmetries, the symmetries are represented as following~\cite{Bi2013}. 
\begin{align}
{\mathbb Z}^{A}_{2} &: z_{e} \rightarrow i\sigma^{z} z_{e}\nonumber\\ 
&: z_{m} \rightarrow z_{m} \nonumber\\
{\mathbb Z}^{B}_{2} &: z_{e} \rightarrow i\sigma^{y} z_{e}\nonumber\\ 
&: z_{m} \rightarrow i\sigma^{z}z_{m}\nonumber\\ 
{\mathbb Z}^{C}_{2} &: z_{e} \rightarrow i\sigma^{z} z_{e}\nonumber\\ 
&: z_{m} \rightarrow \sigma^{x}z_{m} 
\label{symmetry3}
\end{align} 
All the three symmetries are unitary and on-site. The symmetries on the current $J \sim iz^{*}\partial z + h.c.$ \eqref{Current} are represented by the $4 \times 4$ matrices as before. 
\begin{align}
{\mathbb Z}^{A}_{2} &: J^{a}_{\mu} \rightarrow \delta^{ab}J^{b}_{\mu}, \nonumber\\
{\mathbb Z}^{B}_{2} &: J^{a}_{\mu} \rightarrow X^{ab}_{T}J^{b}_{\mu}, \nonumber\\ 
{\mathbb Z}^{C}_{2} &: J^{a}_{\mu} \rightarrow Y^{ab}_{T}J^{b}_{\mu}
\end{align}
\beq
X = 
\left[ 
\begin{array}{cccc} 
0 & -1 & 0 & 0\\
-1 & 0 & 0 & 0\\
0 & 0 & -1 &0\\
0&  0 &0& -1  
\end{array} 
\right] \in GL(4; {\mathbb Z})
\label{op3}
\eeq
\beq
Y = 
\left[ 
\begin{array}{cccc} 
-1 & 0 & 0 & 0\\
0 & -1 & 0 & 0\\
0 & 0 & 0 &-1\\
0&  0 &-1& 0  
\end{array} 
\right] \in GL(4; {\mathbb Z})
\label{op4}
\eeq
Notice that ${\mathbb Z}^{A}_{2}$ is the symmetry of the $K$-matrix \eqref{Kmatrix} and thus does not change the $S$- and $T$- matrices because $\delta^{ab} \in GL(4, {\mathbb Z})$ and $K$ is invariant under the symmetry. On the other hand, ${\mathbb Z}^{B}_{2}$ and ${\mathbb Z}^{C}_{2}$ are not a symmetry of the $K$-matrix but only quantum symmetries. 

Now we proceed to gauge the ${\mathbb Z}^{A}_{2}$ symmetry. As before, we begin with identifying a new anyon excitation $l_{v}$ in the charge lattice~\cite{Lu2013}. 
\beq
{\vec l}_{v} = K \frac{ \delta {\vec \phi}}{2\pi} = (0,0,0,\frac{1}{2})^{T},
\label{deconfined_vortex3}
\eeq
in which we have used $\delta {\vec \phi} = (\pi/2, -\pi/2, 0, 0)^{T}$ from \eqref{symmetry3}. Notice that $X \delta {\vec \phi}$ and $Y \delta {\vec \phi}$ are $\delta {\vec \phi}$ and $-\delta {\vec \phi}$, and thus it is enough to deconfine ${\vec l}_{v}$ above \eqref{deconfined_vortex3} to gauge the ${\mathbb Z}^{A}_{2}$ symmetry. With this new excitation, the allowed anyon excitation in the gauged theory can be represented. 
\beq
{\vec l} = (n_{1},n_{2},n_{3}, n_{4}+\frac{n_{v}}{2})^{T}
\label{AnyonLattice3}
\eeq
with the statistical angle $\theta = \pi l^{T} K^{-1} l $. Notice that the ${\mathbb Z}^{C}_{2}$ symmetry operation $Y$ \eqref{op4} does not leave the anyon lattice \eqref{AnyonLattice3} invariant because it maps $(0,0,0,1/2)^{T}$, the new anyonic excitation, to $(0,0,-1/2,0)^{T}$ which is not allowed in the theory. This already signals that the symmetry ${\mathbb Z}^{C}_{2}$ is not the symmetry of the gauged theory. However ${\mathbb Z}^{B}_{2}$ symmetry operation $X$ {\it does} leave the anyon lattice \eqref{AnyonLattice3} invariant. The effect of ${\mathbb Z}^{B}_{2}$ manifests only when we compute $S$- and $T$- matrices in the gauged theory as we will see soon. 

We extend the anyon lattice with $K$ to the integral anyon lattice with the new $K_{g}$ matrix to describe the theory properly. To rescale the lattice and the $K$-matrix, we find a matrix $M$ such that,
\beq
{\vec l} = (n_{1},n_{2},n_{3}, n_{4}+\frac{n_{v}}{2})^{T} = M {\vec m}, \quad {\vec m} \in {\mathbb Z}^{4}.
\eeq 
We choose $M$ which is bijective.  
\beq
M = \left[ 
\begin{array}{cccc} 
1 & 0 & 0 &0\\
0 & 1 & 0 &0\\
0& 0& 1 & 0\\
0& 0 & 0 &\frac{1}{2}  
\end{array} 
\right] 
\eeq
From $M$, we obtain the new $K_{g}$ matrix for the gauged theory by requiring,  
\beq
M^{T}K^{-1}M = K^{-1}_{g}. 
\eeq

As a result of gauging the symmetry ${\mathbb Z}_{2}$, we end up with the theory, 
\beq
K_{g} = \left[ 
\begin{array}{cccc} 
0 & 0 & 1 &1\\
0 & 0 & 2 &-2\\
1& 2& 0 & 0\\
1& -2 & 0 &0  
\end{array} 
\right],  
\eeq
with the integral anyonic lattice ${\vec m} \in {\mathbb Z}^{4}$. 

We now ask if this gauged $K_{g}$-matrix can generate well-defined $S$- and $T$- matrices invariant under ${\mathbb Z}^{B}_{2} \times {\mathbb Z}^{C}_{2}$ as the case before gauging. The original excitations are represented by $z_{e1} \sim (1,0,0,0)^{T}$, $z_{e2} \sim (0,1,0,0)^{T}$, $z_{m1} \sim (0,0,1,0)^{T}$, and $z_{m2} \sim (0,0,0, 2)^{T}$. From this information, we can easily read off the action ${\tilde X}$ and ${\tilde Y}$ of the symmetries ${\mathbb Z}^{B}_{2}$ and ${\mathbb Z}^{C}_{2}$ as before. 
\beq
{\tilde X} = \left[ 
\begin{array}{cccc} 
0 & -1 & 0 &0\\
-1 & 0 & 0 &0\\
0& 0& -1 & 0\\
0& 0 & 0 & -1 
\end{array} 
\right] 
\eeq
\beq
{\tilde Y} = \left[ 
\begin{array}{cccc} 
-1 & 0 & 0 &0\\
0 & -1 & 0 &0\\
0& 0& 0 & -\frac{1}{2}\\
0& 0 & -2 & 0 
\end{array} 
\right] 
\eeq
We notice that ${\tilde Y}$ is not an element of $GL(4, {\mathbb Z})$ and does not leave the anyon lattice invariant. Hence ${\tilde S}$- and ${\tilde T}$- matrices \eqref{TopMat} defined by $K_{g}$ are not invariant under the actions of ${\tilde Y}$. This shows that the symmetry ${\mathbb Z}^{C}_{2}$ is broken if we gauge the symmetry ${\mathbb Z}^{A}_{2}$. On the other hand, ${\tilde X}$ {\it is} an element of $GL(4, {\mathbb Z})$ and {\it does} leave the anyon lattice invariant. However, ${\tilde X}$ does not leave ${\tilde S}$- and ${\tilde T}$- matrices invariant, {\it i.e.} ${\tilde X}^{T} K^{-1}_{g} {\tilde X} = K^{-1}_{g}$ mod $N$ where $N$ is a matrix with the integers ${\mathbb Z}$ for the off-diagonal elements and $2{\mathbb Z}$ for the diagonal elements. Thus we conclude that ${\mathbb Z}^{B}_{2}$ is also broken in the gauged theory.

\section{Conclusion}

In this paper, we have considered the theory for the topologically ordered surface states of three dimensional bosonic SPT phases with the discrete symmetries $G_{1} \times G_{2}$,
in which one of the symmetries is a conventional symmetry. 
Then we have demonstrated that the multi-component Chern-Simons effective theory itself is not invariant under the symmetry transformations, but the topological $S$- and $T$- matrices defined by the Chern-Simons theory are invariant under the symmetries.

However, if we gauge one of the symmetries $G_{1} \times G_{2}$ on the surface by pretending that the surface is a purely two-dimesnional system, we find that the other symmetry must be broken in the gauged theory because the topological $S$- and $T$- matrices defined by the Chern-Simons theory for the gauged surface are not invariant under the other symmetry. 
This reminds us of the edge theory of the two-dimensional SPT phases 
in which we also find the conflict of the symmetries. 
The conflict of the symmetries at the edge of the two-dimensional SPT phases signals that the edge theory can never be realized as a purely one-dimensional lattice model. 
Given the fact that there has been no such phenomena in purely two-dimensional topologically ordered states, this signals that the symmetries are encoded ``anomalously'' on the surface of the SPT phases and that the surface state cannot be realized on the purely two-dimensional lattice models. Presumably, if gauging the symmetry is done in bulk as a whole, then the ``anomaly'' or ``conflict of symmetries'' on the surface is expected to be cancelled by certain contributions from bulk. At this point, it is not clear how to see this from this work, and so we leave this as the future problem.  

{\it Note added}: Upon compeletion of this work, we became aware of a paper by Kapustin et.al.~\cite{Kapustin} and a unpublished work~\cite{ChengUnpublished}. In the reference~\cite{Kapustin}, the authors also considered the SPT phases with the discrete $G \times H$ and found the breaking of the symmetry $H$ upon gauging the symmetry $G$. In the reference~\cite{ChengUnpublished}, the authors found an obstruction to guage $G_{1}\times G_{2}$ on the surface of the three-dimensional bosonic SPT phases. 

\acknowledgements
Authors thank Chetan Nayak and Meng Cheng for helpful discussion. The authors acknowledge support from NSF grant DMR-1064319 and ICMT postdoctoral fellowship (G.Y.C.), and Simons foundation (J.C.Y.T). 

\bibliography{Melting}

\end{document}